\documentclass[a4paper,floatfix,aps,pra,showpacs,twocolumn,amsmath,preprintnumbers,nofootinbib,secnumarabic,superscriptaddress]{revtex4}

\usepackage{epsfig,amssymb,graphicx}
\usepackage{color}

\definecolor{newgreen}{rgb}{0,.5,0}

\def\scn#1#2{\section{#1}\lb{#2}}
\def\sscn#1#2{\subsection{#1}\lb{#2}}
\def\bfl{\begin{flushleft}}
\def\efl{\end{flushleft}}
\def\bfr{\begin{flushright}}
\def\efr{\end{flushright}}
\def\bc{\begin{center}}
\def\ec{\end{center}}
\def\be{\begin{equation}}
\def\ee{\end{equation}}
\def\ba{\begin{eqnarray}}
\def\ea{\end{eqnarray}}
\def\baa#1{\begin{array}{#1}}
\def\eaa{\end{array}}
\def\bw{\begin{widetext}}
\def\ew{\end{widetext}}
\def\nn{\nonumber }
\def\lb#1{\label{#1}}
\def\bit{\begin{itemize}}
\def\eit{\end{itemize}}
\def\bco{\begin{comment}} 
\def\eco{\end{comment}}

\def\twomat#1#2#3#4{\begin{pmatrix} #1 & #2 \\ #3 & #4 \end{pmatrix}}
\def\twomatsm#1#2#3#4{\left(\begin{smallmatrix} #1 & #2 \\ #3 & #4 \end{smallmatrix}\right)}

\def\Sinh#1#2{\, \text{sinh}^{#1}\!\left(#2 \right) }

\def\densnnorm{\hat{\Omega}}

\begin{document}

\preprint{\small J. Stat. Mech. (2016) 033102 [arXiv:1502.07086]}

\title{
Quantum entropy of systems  described by non-Hermitian Hamiltonians
}

\author{Alessandro Sergi}
\email{asergi@unime.it}

\affiliation{
Dipartimento di Fisica e Scienze della Terra,
Universit\'a degli Studi di Messina,
Contrada Papardo, 98166 Messina, Italy
}

\affiliation{
School of Chemistry and Physics, University of KwaZulu-Natal, 
Private Bag X01, Scottsville, Pietermaritzburg 3209, South Africa}

\author{Konstantin G. Zloshchastiev}
\email{k.g.zloschastiev@gmail.com}
\affiliation{
Institute of Systems Science, Durban University of Technology, P. O. Box 1334, Durban 4000, South Africa}

%\date\today
\date{\footnotesize Received: 25 February 2015 [arXiv], 1 September 2015 [JStat]}

\begin{abstract}
We study the quantum entropy of systems that are described by
general non-Hermitian Hamiltonians, including those which can model the
effects of sinks or sources. We generalize the von Neumann entropy to the non-
Hermitian case and find that one needs both the normalized and non-normalized
density operators in order to properly describe irreversible processes. It turns out
that such a generalization monitors the onset of disorder in quantum dissipative
systems. We give arguments for why one can consider the generalized entropy
as the informational entropy describing the flow of information between the
system and the bath. We illustrate the theory by explicitly studying few simple
models, including tunneling systems with two energy levels and non-Hermitian detuning.
\end{abstract}


\maketitle

\section{Introduction}

One of the most intriguing problems of statistical mechanics
is provided by the fact that Hamiltonian reversible dynamics is not able
to predict any increase of the fine-grained entropy, as it would be required by 
the second law of thermodynamics \cite{callen,aharonov,balescu}.
However, it has been shown that for classical systems
such an increase can be described through
the adoption of non-Hamiltonian dynamics~\cite{nose,hoover,b,b2,aspvg}
with phase space compressibility~\cite{andrey,andrey2}.

The difficulties with the reconciliation of the fine-grained entropy and
thermodynamics remain unchanged when passing to the realm of quantum mechanics.
Here, we consider the quantum dynamics originating
from general non-Hermitian Hamiltonians (NH), known as the non-Hermitian approach.
This approach is often invoked in order to describe quantum systems coupled
to sinks or sources and it may arise in a variety of contexts, 
for instance, when studying
optical waveguides \cite{optics,optics2},
Feshbach resonances and
particles' disintegration \cite{nimrod2,seba,spyros,fesh,fesh2,sudarshan},
multiphoton ionization \cite{selsto,baker,baker2,chu},
and open quantum systems \cite{kor64,wong67,heg93,bas93,ang95,rotter,rotter2,gsz08,bellomo,banerjee,reiter,bg12,sz14,ks}.
In all such cases, the probability 
does not have to be conserved, in general.

In this work, we
show that the production of the 
fine-grained entropy
can be
naturally predicted within the framework of the
non-Hermitian approach.
In particular,
we extend the definition of the
Gibbs-von Neumann entropy~\cite{vonneumann}
to the case of systems with non-Hermitian Hamiltonians
and introduce a ``non-Hermitian'' entropy combining
the normalized and non-normalized density matrix. 
%We show
%that, contrary to the Hermitian case, the rate of entropy production
%is not identically zero.
In order to illustrate the theory,
we explicitly consider the analytical solution of some models
of interest for quantum dynamics.
Depending on the model studied,
we find that the non-Hermitian entropy can provide the expected
behavior at large times. 
%However, we also find cases where this does not occur and cases in which the entropy becomes negative. 

The structure of this paper is as follows.
In Sec. \ref{s-nh} we give a brief outline of the
density operator approach for NH systems.
In Sec. \ref{s-ent} we introduce a generalization
of the Gibbs-von-Neumann entropy that is suitable
for NH systems, and discuss its features.
In Sec. \ref{s-ex} we study the NH dynamics of a two-level system
in order to illustrate the formalism.
The discussion of the results and the conclusions are presented in Sec. \ref{s-con}.

\scn{Quantum dynamics with non-Hermitian Hamiltonians}{s-nh}

In the theory of open quantum systems the non-Hermitian approach
has recently acquired a 
strong popularity since it has a different range of applicability 
from the approach based on the Lindblad master equation \cite{bpbook}. 
In order to sketch how the approach unfolds, one can consider
a total non-Hermitian Hamiltonian
\begin{equation}
\hat{\cal H}=
\hat H -i \hat\Gamma \;,
\label{eq:tot-NH-H}
\end{equation}
where both $\hat H$ and $\hat\Gamma$ are Hermitian 
($\hat\Gamma$ is often called the decay rate operator),
the Schr\"odinger equations for the quantum states $|\Psi\rangle$
and $\langle\Psi|$ are written as
\begin{eqnarray}
\partial_t|{\Psi}\rangle&=&-\frac{i}{\hbar}\hat{\cal H}|\Psi\rangle=
-\frac{i}{\hbar}\hat H|\Psi\rangle-\frac{1}{\hbar}\hat\Gamma|\Psi\rangle
\;, \\
\partial_t \langle {\Psi}|&=&\frac{i}{\hbar}\langle\Psi |\hat{\cal H}^\dagger
=\frac{i}{\hbar}\langle\Psi|\hat H-\frac{1}{\hbar}\langle\Psi|\hat\Gamma
.
\end{eqnarray}
Upon introducing a non-normalized density matrix 
\begin{equation}
\hat{\Omega} =\sum_k{\cal P}_k|\Psi^k \rangle\langle\Psi^k |
\;,
\end{equation}
where ${\cal P}_k$ are the probabilities of the states 
$(|\Psi^k \rangle,\langle\Psi^k |)$ that are compatible with
the macroscopic constraints obeyed by the system,
the dynamics can be recast in terms of the equation
\begin{equation}
\partial_t {\hat{\Omega}} =-\frac{i}{\hbar}\left[\hat H, \hat{\Omega}\right]-
\frac{1}{\hbar}
\left\{\hat{\Gamma},\hat{\Omega} \right\}
\;,
\label{eq:dotOmega}
\end{equation}
where $[ , ]$ and $\left\{ , \right\}$ denote the commutator and anticommutator,
respectively.
In the context of theory of open quantum systems, the evolution equation
for the density operator $\densnnorm$
effectively describes the original 
subsystem (with Hamiltonian $\hat{H}$) together with the effect of environment (represented
by $\hat{\Gamma}$). 

Upon taking the trace of both sides of Eq.~(\ref{eq:dotOmega}),
one obtains an evolution equation for the trace of $\hat{\Omega} $:
\begin{equation}
\partial_t{\rm Tr}\,\hat{\Omega}
=-\frac{2}{\hbar}{\rm Tr}\left(\hat{\Gamma} \,\hat{\Omega} \right)
\;.
\label{eq:dotTrOmega}
\end{equation}
Equation~(\ref{eq:dotTrOmega}) shows that NH dynamics does not conserve
the probability.

As suggested in Ref.~\cite{ks}, one is then led to the introduction
of a normalized density matrix, defined as
\begin{equation}
\hat{\rho} =\frac{\hat{\Omega} }{{\rm Tr}\,\hat{\Omega} } \;,
\label{eq:rho}
\end{equation}
that can be used in the calculation of quantum statistical averages
of arbitrary operators $\hat\chi$:
$
\langle\chi\rangle_t={\rm Tr} \left( \hat\chi\hat{\rho} \right)  .
$

As a result of the definition given in Eq.~(\ref{eq:rho}) and the evolution
equations in~(\ref{eq:dotOmega}) and~(\ref{eq:dotTrOmega}), the normalized density
matrix obeys a dynamics ruled by the following equation:
\begin{equation}
\partial_t {\hat{\rho}} =
-\frac{i}{\hbar}\left[\hat H, \hat{\rho}\right]
-\frac{1}{\hbar}\left\{\hat{\Gamma},\hat{\rho} \right\}
+\frac{2}{\hbar}\hat{\rho} \, {\rm Tr} (\hat{\Gamma}\hat{\rho})
.
\label{eq:dotrho}
\end{equation}
This equation effectively describes the evolution of 
original subsystem (with Hamiltonian $\hat{H}$) together with the effect of 
environment (represented
by $\hat{\Gamma}$)
and the additional term that restores the 
overall probability's conservation.  
One can see that due to the last term in this equation, the dynamics
of the normalized density matrix $\hat{\rho}$ is nonlinear.
A similar nonlinearity was found in the
evolution equation for the operator averages \cite{emg-korsch}.
Moreover, the appearance of nonlinearities in NH-related theories has
also been suggested in Ref. \cite{zno02},
on the grounds of the Feshbach-Fano projection formalism.

The density operator $\hat{\rho}$,
determined by the solution of Eq. (\ref{eq:dotrho}),
is bounded and
allows one to maintain a probabilistic interpretation
of the statistical averages of operators under non-Hermitian dynamics.
Nevertheless, the gain or loss of probability associated with
the coupling to sinks or sources are properly described
by the non-normalized density operator $\hat{\Omega}$.
Hence, it turns out that one must use both $\hat{\Omega}$ and $\hat{\rho}$
in the formalism, one to describe gain or loss of probability and
the other to calculate averages.
We have already verified in 
our previous work \cite{sz14cor} on time correlation functions
the need to consider both $\hat{\Omega}$ and $\hat{\rho}$
in the definitions of statistical properties.

\scn{Quantum entropy}{s-ent}

It is well-known that in the Hermitian case the  quantum dynamics is unitary
and defined 
in terms of a normalized
density matrix $\hat{\Xi} $ obeying the quantum Liouville equation of motion:
\begin{equation}
\partial_t {\hat{\Xi}} =-\frac{i}{\hbar}\left[\hat H, \hat{\Xi}\right]
.
\label{eq:dotXi}
\end{equation}
The quantum entropy can be defined as 
\begin{equation}
S_{\rm vN}
\equiv
-k_B{\rm Tr} (\hat{\Xi} \ln \hat{\Xi})
,
\label{eq:SVNH}
\end{equation}
where $k_B$ is Boltzmann's constant \cite{vonneumann}.
The rate of entropy production, derived from the quantum Liouville 
equation in~(\ref{eq:dotXi}), is
\begin{equation}
\partial_t {S}_{\rm vN}
=
-k_B{\rm Tr}\left(\partial_t {\hat{\Xi}}  \, \ln\hat{\Xi} \right)
= 0
.
\label{eq:dotSVNH}
\end{equation}
While the von Neumann entropy in Eq.~(\ref{eq:SVNH}) is fit to represent
the properties of equilibrium quantum systems,
Eq.~(\ref{eq:dotSVNH}) implies that the use of the entropy in Eq.~(\ref{eq:SVNH})
is somewhat more problematic in nonequilibrium dynamics.
In fact, in order to agree with the entropy
increase required by the second law of thermodynamics, one must
resort to modified definitions of entropy, such as those implied by coarse-graining 
(see Ref.~\cite{oerter},
for example) or by the adoption of relevant definitions of entropy \cite{balian}.
This has even led some authors~\cite{andrey,andrey2}
to invoke more general structures~\cite{nose,hoover,b,b2,aspvg}
than Hamiltonian ones in order to
define the microscopic dynamics of statistical systems.

When a quantum system is coupled to sink or sources, NH dynamics can be used.
In such a case the straightforward adoption of the von Neumann entropy~(\ref{eq:SVNH})
leads to
\begin{equation}
S_{\rm vN}
\equiv
-k_B \left\langle   \ln\hat{\rho} \right\rangle
=
-k_B{\rm Tr}\left( \hat{\rho} \ln\hat{\rho} \right)
.
\label{eq:SVNnH}
\end{equation}
Equation~(\ref{eq:SVNnH}) clearly becomes identical to
Eq.~(\ref{eq:SVNH}) when $\hat{\Gamma}\to 0$ so that
$\hat{\cal H} \to \hat H$ and the dynamics becomes unitary.

If one takes the time derivative of Eq.~(\ref{eq:SVNnH}),
uses the evolution equation in~(\ref{eq:dotrho})
and the properties of the trace, the following equation
for the rate of entropy production is obtained:
\begin{eqnarray}
\partial_t S_{\rm vN}
= 
\frac{2k_B}{\hbar}{\rm Tr}\left(\hat{\Gamma}\hat{\rho} \ln\hat{\rho} \right)
+
\frac{2}{\hbar}{\rm Tr}\left( \hat{\Gamma}\hat{\rho} \right)
S_{\rm vN}
. \label{eq:dotSVNnH}
\end{eqnarray}
This equation shows that the non-unitary evolution
given by Eq.~(\ref{eq:dotrho}) provides, in general, a non-zero
entropy production. 

Interestingly, in agreement with our discussion about the important role
of both $\hat{\Omega}$ and $\hat{\rho}$ in non-Hermitian dynamics, done in
Sec.~\ref{s-nh}, it is also possible to define the entropy
as the statistical average of the logarithm
of the non-normalized density operator:
\begin{equation}
S_\text{NH}
\equiv
-k_B \langle   \ln\hat{\Omega} \rangle
=
-k_B{\rm Tr} ( \hat{\rho} \ln\hat{\Omega} ) 
= 
- k_B 
\frac{
{\rm Tr}( \hat{\Omega} \ln\hat{\Omega} )
     }{
{\rm Tr}\, \hat{\Omega}
}
,
\label{eq:SVNnH2}
\end{equation}
with the evolution of $\hat{\Omega}$
naturally given by the linear equation (\ref{eq:dotOmega}).
One can expect $S_{\rm vN}$ not to be able to catch properly 
the gain or loss of probability
because of its sole reliance on the bounded $\hat{\rho}$ with its
nonlinear corrections.
Instead, the operator $\ln \hat{\Omega}$
can be expected to monitor properly the probability evolution.
The rate of change of $S_\text{NH}$ is easily found to be
\begin{equation}
\partial_t S_{\rm NH}
=\frac{2k_{\rm B}}{\hbar}{\rm Tr}\left(\hat{\Gamma}\hat{\rho}\ln\hat{\Omega}\right)
+\frac{2}{\hbar} {\rm Tr}\left(\hat\Gamma \rho\right) S_{\rm NH}
+2\frac{k_B}{\hbar}{\rm Tr}\left(\hat{\Gamma}\hat{\rho}\right)\;.
\end{equation}
The two entropies are related by the formula
\be\lb{e:diffentr}
S_\text{NH}
=
S_{\rm vN}
-
k_{\rm B}
\ln\left( \text{Tr}\, \hat\Omega\right)
,
\ee
therefore, the difference between $S_\text{NH}$ and $S_{\rm vN}$ is a measure of deviation of $\text{Tr}\, \hat\Omega$
from unity. 

Another important property of the entropy $S_\text{NH}$ is that,
unlike the von Neumann entropy (\ref{eq:SVNnH}),
it is not invariant under 
the complex constant shifts of 
the Hamiltonian
that preserve the form of the evolution equation 
for the normalized density operator (\ref{eq:dotrho}). 
These constant shifts of the Hamiltonian
can be regarded as a kind of ``gauge'' transformation,
see Appendix.
%see proof in the Appendix.
Indeed, if one adds to the 
$\hat\Gamma$ operator a constant term
that is proportional to the unity operator
$\hat I$, then both the normalized density operator (\ref{eq:rho})
and the von Neumann entropy (\ref{eq:SVNnH}) 
are unchanged;  however, the NH entropy acquires a shift
in terms of a linear function of time:
\be\lb{eq:gaugetrans}
\hat \Gamma \to 
\hat \Gamma + 
\frac{1}{2}
\hbar
\alpha
\hat I
\ 
\Rightarrow
\ 
\left\{
\baa{l}
S_\text{vN} \to S_\text{vN} ,
\\
S_\text{NH} \to S_\text{NH} + k_B \alpha t
,
\eaa
\right.
\ee
where $\alpha$ is an arbitrary real constant, see Appendix for details.
Such a property
can facilitate the computing of $S_\text{NH}$ for some systems.
It also throws light on the fact that
the NH entropy ``remembers'' the effects
of the complex constant shifts,
$\hat{\cal H} \to \hat{\cal H} + c_0 \hat I$ (where $c_0$ is
an arbitrary complex number).
%, eventually performed on the non-Hermitian Hamiltonian.

\scn{Examples}{s-ex}

In order to demonstrate the behavior of 
the above-mentioned types of entropy, in this
section we consider few simple models.

\sscn{Models with a constant $\hat\Gamma$ operator}{s-gau}

This is the class of models where the Hermitian part of the Hamiltonian
can be any physically admissible self-adjoint
operator $\hat H$
whereas the $\hat\Gamma$ operator is proportional to the 
identity operator:
\begin{equation}
\hat\Gamma
=
\frac{1}{2}
\hbar
\gamma_0 
\hat I
,
\label{eq:Gammamodgau}
\end{equation}
where the parameter 
$\gamma_0$ is assumed to be real-valued.

In such models the value
of the parameter $\gamma_0 $ 
does not affect the time evolution of the
normalized density operator (\ref{eq:rho}).
Indeed, Eq. (\ref{eq:dotrho})
becomes just the conventional quantum Liouville equation:
\begin{equation}
\partial_t {\hat{\rho}} =
-\frac{i}{\hbar}\left[\hat H, \hat{\rho}\right]
%-\frac{1}{\hbar}\left\{\hat{\Gamma},\hat{\rho} \right\}+\frac{2}{\hbar}\hat{\rho} \, {\rm Tr} (\hat{\Gamma}\hat{\rho})
.
\label{eq:dotrhogau}
\end{equation} 
However, 
equation (\ref{eq:dotOmega}) reveals that
the operator (\ref{eq:Gammamodgau}) does affect the evolution of 
the operator $\hat\Omega$ and of the entropy $S_{\rm NH}$.
Equations (\ref{eq:dotTrOmega})
and (\ref{eq:dotSVNnH}) become, respectively:
\ba
&&
\partial_t{\rm Tr}\,\hat{\Omega}
=
%-\frac{2}{\hbar}{\rm Tr}\left(\hat{\Gamma} \,\hat{\Omega} \right)
-\gamma_0 {\rm Tr}\,\hat{\Omega}
,\\&&
\partial_t S_{\rm vN}
%= \frac{2k_B}{\hbar}{\rm Tr}\left(\hat{\Gamma}\hat{\rho} \ln\hat{\rho} \right) + \frac{2}{\hbar}{\rm Tr}\left( \hat{\Gamma}\hat{\rho} \right) S_{\rm vN}
=
k_B \gamma_0 {\rm Tr}\left(\hat{\rho} \ln\hat{\rho} \right)
+
\gamma_0 
%{\rm Tr}\left( \hat{\rho} \right)
S_{\rm vN}
\equiv
0
.
\ea
Imposing the initial conditions ${\rm Tr}\,\hat{\Omega} (0) = 1$
and $S_{\rm vN} (0) = S_{\rm vN}^{(0)} = \text{const}$,
and also using the relation (\ref{e:diffentr}),
we obtain
\ba
&&
{\rm Tr}\,\hat{\Omega} (t) = \exp{(-\gamma_0 t)}
,\\&&
S_{\rm vN} (t) = S_{\rm vN}^{(0)} = \text{const}
,\\&&
S_\text{NH} (t)
% = S_{\rm vN} (t) - k_{\rm B} \ln\left( \text{Tr} \hat\Omega (t)\right)
=
S_{\rm vN}^{(0)} + k_{\rm B} \gamma_0 t
.
\ea
One can see that at large times the trace of $\hat{\Omega}$ either
diverges (at negative values of $\gamma_0$) or vanishes
(at positive values of $\gamma_0$) 
but the conventional von Neumann entropy $S_{\rm vN}$ does not
reflect this behaviour in any way.
On the other hand,  the entropy $S_\text{NH}$ 
provides more information in this regard.
For instance, at positive values of $\gamma_0$ the trace of 
$\hat\Omega$ goes asymptotically to zero,
describing the damping of the probability.
In such a case,
$S_\text{NH}$ grows linearly with time as the thermodynamic entropy is expected
to do.
Further discussion of these
features is given in the concluding section.

\sscn{Two-level tunneling model with non-Hermitian detuning}{s-tldet}

Let us consider a two-level model specified by the Hermitian Hamiltonian
\begin{equation}
\hat H=-\hbar \Delta \hat\sigma_x
\label{eq:Hmod}
\end{equation}
and the $\hat\Gamma$ operator 
%(up to a gauge transformation (\ref{eq:gaugetrans}))
\begin{equation}
\hat\Gamma
=
\hbar
\gamma 
\hat\sigma_z 
,
\label{eq:Gammamod}
\end{equation}
with the total Hamiltonian given by Eq.~(\ref{eq:tot-NH-H}).
The parameters $\Delta$ and $\gamma$ are real-valued,
with $\Delta$ being also positive.
The symbols $\hat\sigma_x$ and $\hat\sigma_z$ denote the Pauli matrices:
$
\hat\sigma_x = \twomatsm 0 1 1 0 $,
$
\hat\sigma_z = \twomatsm{1}{0}{0}{-1}
$.
This model is the non-Hermitian analogue of the well-known tunneling model
with detuning \cite{leggett}, which finds applications in the pseudo-Hermitian and
$PT$-symmetric quantum mechanic \cite{sgh92,ben98}.
Such kind of models
are often used in order to effectively describe
the dissipative and measurement-related phenomena
in open quantum-optical 
and spin systems, such as the direct photodetection of a driven TLS 
interacting with the electromagnetic field \cite{bpbook}.

\begin{figure}[htbt]
\begin{center}\epsfig{figure=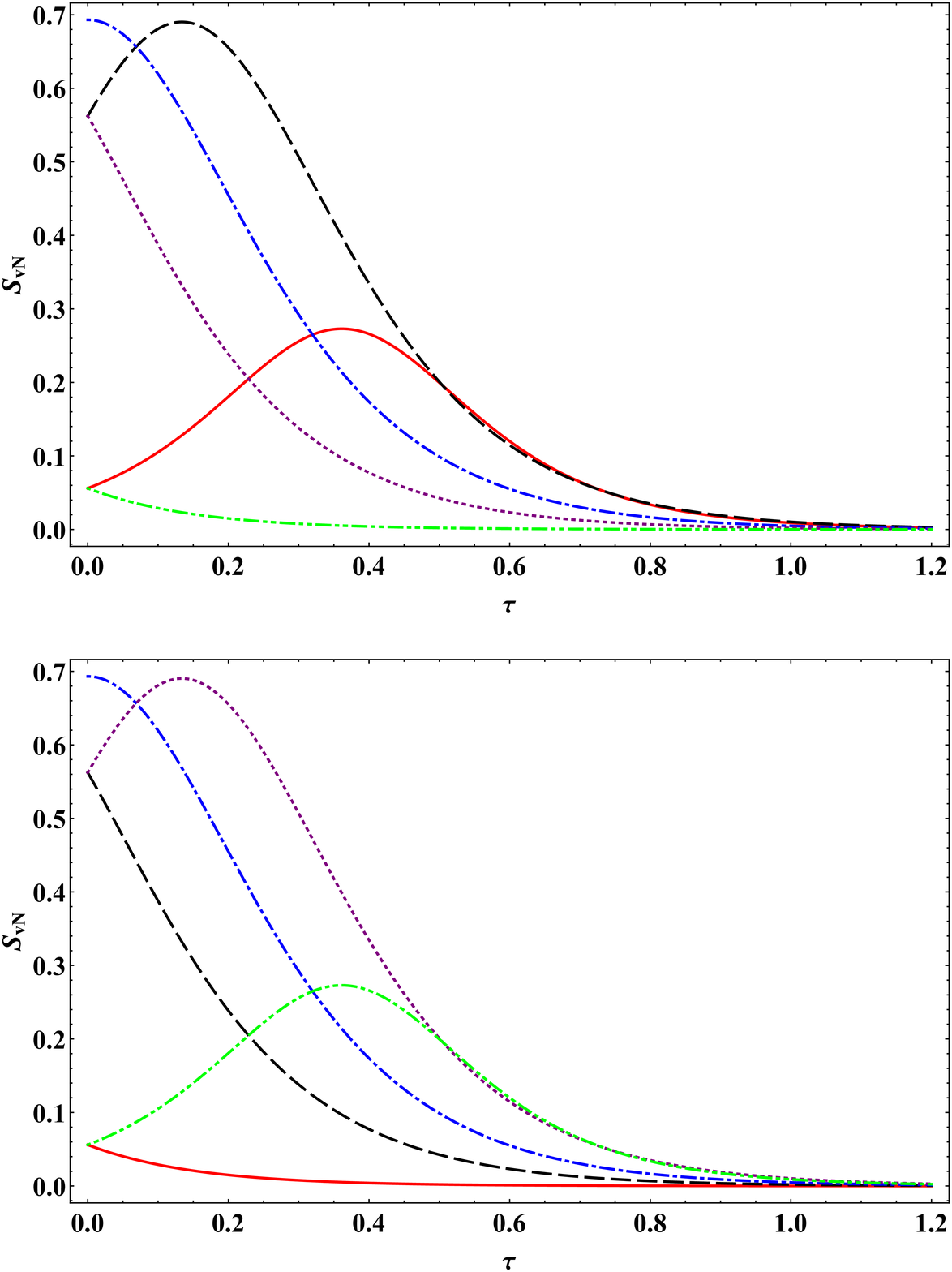,width=  0.99\columnwidth}\end{center}
\caption{Entropy $S_{\rm vN}$ (\ref{e:m1entr}) versus $\tau$
for the values of  $\tilde\gamma = -2$ (top plot)
and $\tilde\gamma =  2$ (bottom plot),
at the different values of $p$:
$0.01$ (solid line),
$1/4$ (dashed line),
$1/2 $ (dash-dotted line),
$3/4 $ (dotted line)
and
$0.99$ (dash-double-dotted line).}
\label{f-m1-gampm2}
\end{figure}

As an initial state we choose the superposition of the ground and
excited states
\be\lb{eq:rhoini}
\hat{\rho} (0) =
\hat{\Omega} (0) =
p \left|e\right\rangle \left\langle  e \right|
+
(1-p) \left|g\right\rangle \left\langle  g \right|
=
\twomat{p}{0}{0}{1-p}
,
\ee
where $0 \leqslant p \leqslant 1$ is a free parameter.
It is easy to check that this state is pure at $p=0, \, 1$ and mixed otherwise.
Solving the evolution equations (\ref{eq:dotOmega}),
with the initial condition (\ref{eq:rhoini}),
we obtain the following expression for the non-normalized density operator
\be\lb{e:m1omega}
\hat\Omega 
=
\frac{1}{2 \mu^2}
\left[
2 f_y (t) \hat\sigma_y
+
f_z (t) 
\hat\sigma_z
+
F(t)
\hat I
\right]
,
\ee
where 
%$\hat I$ is the identity matrix, and 
we denoted:
\ba
&&
f_y (t) =
\sinh{(\mu \tau)}
\left[
\mu \bar p  \cosh{(\mu \tau)}
-
\tilde\gamma \sinh{(\mu \tau)}
\right]
,
\\&&
f_z (t) =
\mu
\left[
\mu \bar p  \cosh{(2 \mu \tau)}
-
\tilde\gamma \sinh{(2 \mu \tau)}
\right]
,
\\&&
F (t) 
%= \mu^2 \text{Tr} \hat\Omega
=
\tilde\gamma^2 \cosh{(2 \mu \tau)}
- 
\mu \bar p \tilde\gamma  \sinh{(2 \mu \tau)}
-
1
,
\lb{e:m1funF}
\ea
where $\tau = \Delta t$, $\tilde\gamma = \gamma/ \Delta$,
$\bar p = 2 p -1$, and the value $\mu = \sqrt{\tilde\gamma^2 -1}$ is assumed
to be positive throughout the paper.
One can see that for the chosen initial state (\ref{eq:rhoini}), $\text{Tr}\, \hat\Omega$
is invariant under the simultaneous transformation $\tilde\gamma \to - \tilde\gamma$
and $p \to 1-p$, which will manifest itself in the behavior of $S_\text{NH}$ below.
Consequently, the normalized density matrix is given by:
\be
\hat\rho  
=
\frac{f_y (t)}{F(t)} \hat\sigma_y
+
\frac{f_z (t) }{2 F(t)}
\hat\sigma_z
+ 
\frac{1}{2}
\hat I
.
\ee

The von Neumann entropy can be computed directly from the definition (\ref{eq:SVNnH}).
It is given by (in units where $k_B = 1$):
\be
S_{\rm vN}=
- 
F_+^{(2)} (t)
\,
\text{ln}
F_+^{(1)}(t)
- 
F_-^{(2)}(t)
\,
\text{ln}
F_-^{(1)}(t)
,
\lb{e:m1entr}
\ee
where we denoted
$
F_\pm^{(i)}(t)
=
\tfrac{1}{2}
\left[
1 \pm \sqrt{F_i (t)}/F(t)
\right]
$, 
$i=1, 2$,
and
\bw
\ba
F_1 (t) &=&
4 \Sinh{2}{\mu \tau}
\left[
\mu \bar p  \cosh{(\mu \tau)}
-
\tilde\gamma \sinh{(\mu \tau)}
\right]^2
%\nn\\&&
+
\mu^2
\left[
\mu \bar p  \cosh{(2 \mu \tau)}
-
\tilde\gamma \sinh{(2 \mu \tau)}
\right]^2
,\\
F_2 (t) &=&
\mu \bar p \tilde\gamma
\left[
2 \sinh{(2 \mu \tau)} - \tilde\gamma^2 \sinh{(4 \mu \tau)}
\right] 
%\nn\\&&
+
\frac{1}{2}
\tilde\gamma^2
\left[
(\mu^2 (\bar p^2 +1) +1)
\cosh{(4 \mu \tau)} 
- 
4 \cosh{(2 \mu \tau)}
\right] 
\nn\\&&
+
\frac{1}{2}
\left[
\mu^4 (\bar p^2 -1)
-
\mu^2 (\bar p^2 -2)
+ 3
\right]
.
\ea
\ew
The typical profiles of the entropy (\ref{e:m1entr})
for the initial state (\ref{eq:rhoini})
are shown in Fig. \ref{f-m1-gampm2}.
One can see that 
the entropy $S_{\rm vN}$ tends to zero at large times,
regardless of the sign of $\tilde\gamma$.

The NH entropy, defined in Eq. (\ref{eq:SVNnH2}), can be computed
using the relation (\ref{e:diffentr}).
It turns out to be (in units where $k_B = 1$):
\be
S_\text{NH}
=
S_{\rm vN}
-
\ln{\left[F (t)/\mu^2\right]}
,
\lb{e:m1entr2}
\ee
where $S_{\rm vN}$ is given by (\ref{e:m1entr}).
One can see that if the von Neumann entropy remains finite 
at large times 
then the asymptotical properties of $S_\text{NH}$ are determined
by the behavior of $F(t)$, i.e., 
\be
\lim\limits_{t\to\infty} S_\text{NH} = \ln{\mu^2} - \lim\limits_{t\to\infty} \ln{F (t)}
\propto - 2 \mu \tau
,
\ee
such that $S_\text{NH}$ tends to a linear function at large times.

\begin{figure}[htbt]
\begin{center}\epsfig{figure=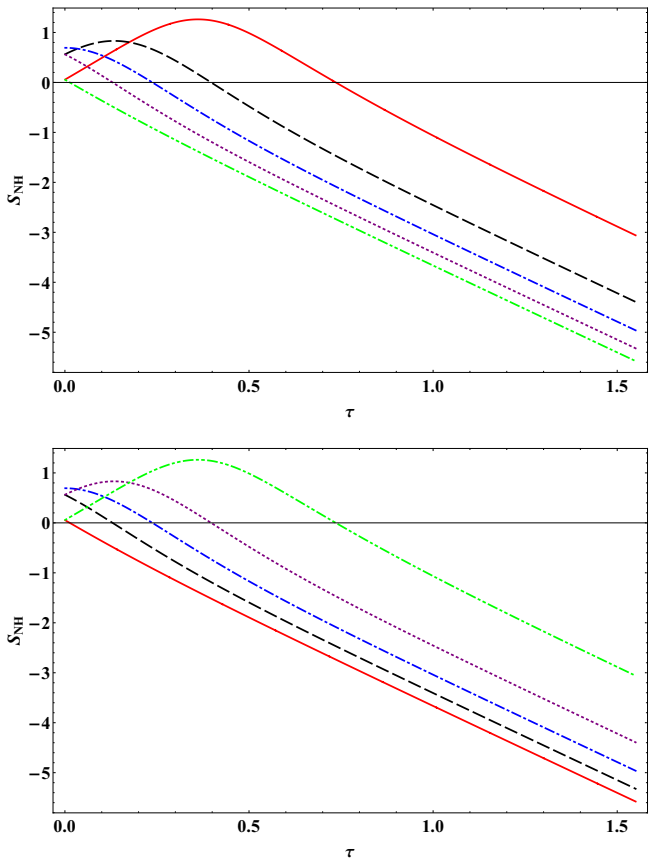,width=  0.99\columnwidth}\end{center}
\caption{Entropy $S_\text{NH}$ (\ref{e:m1entr2}) versus $\tau$
for the values of  $\tilde\gamma = -2$ (top plot)
and $\tilde\gamma =  2$ (bottom plot),
at the different values of $p$:
$0.01$ (solid line),
$1/4$ (dashed line),
$1/2 $ (dash-dotted line),
$3/4 $ (dotted line)
and
$0.99$ (dash-double-dotted line).}
\label{f-m1e2-gampm2}
\end{figure}

\begin{figure}[htbt]
\begin{center}\epsfig{figure=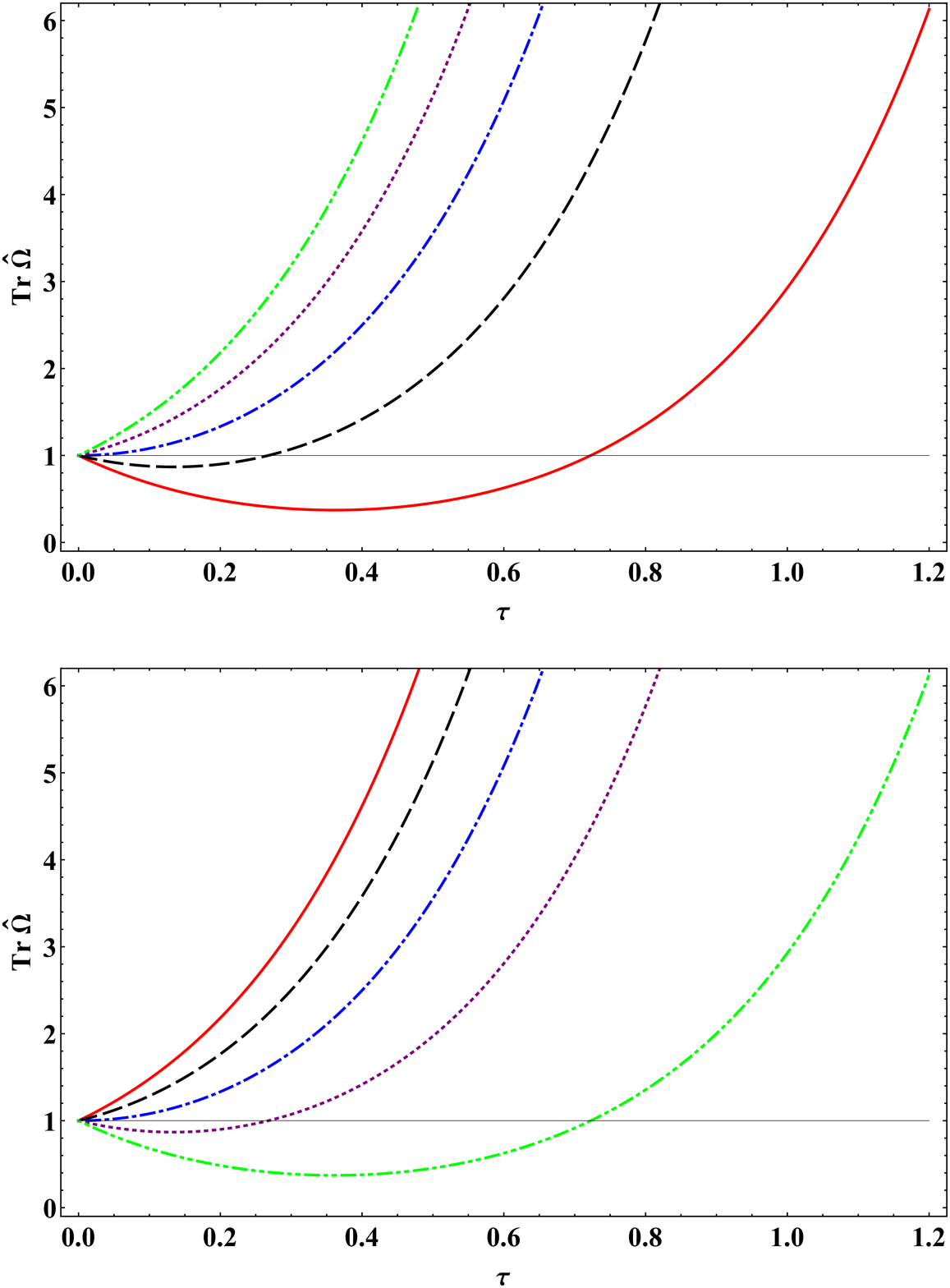,width=  0.99\columnwidth}\end{center}
\caption{Profiles of 
$\text{Tr}\, \hat\Omega$, where $\hat\Omega$ is given by (\ref{e:m1omega}), versus $\tau$
for the values of  $\tilde\gamma = -2$ (top plot)
and $\tilde\gamma =  2$ (bottom plot),
at the different values of $p$:
$0.01$ (solid line),
$1/4$ (dashed line),
$1/2 $ (dash-dotted line),
$3/4 $ (dotted line)
and
$0.99$ (dash-double-dotted line).}
\label{f-m1tr-gampm2}
\end{figure}

The profiles of the entropy (\ref{e:m1entr2})
%and (\ref{e:m1entr2cal})
for the initial state (\ref{eq:rhoini})
are shown in Fig. \ref{f-m1e2-gampm2}.
One can see that 
the entropy decreases with time, regardless
of the sign of $\tilde\gamma$, as one can expect from equations (\ref{e:m1funF}) and (\ref{e:m1entr2}).
It is instructive to compare this with the plots
for the trace of the non-normalized density matrix,
shown in Fig. \ref{f-m1tr-gampm2}, which indicate
the flow rate of probability to/from the system.
For this model, the entropy $S_\text{NH}$
takes negative values at large times.
This will be discussed in details in the concluding section.

\sscn{Two-level tunneling model with non-Hermitian detuning and asymptotically constant NH entropy}{s-tldet2}

\begin{figure}[htbt]
\begin{center}\epsfig{figure=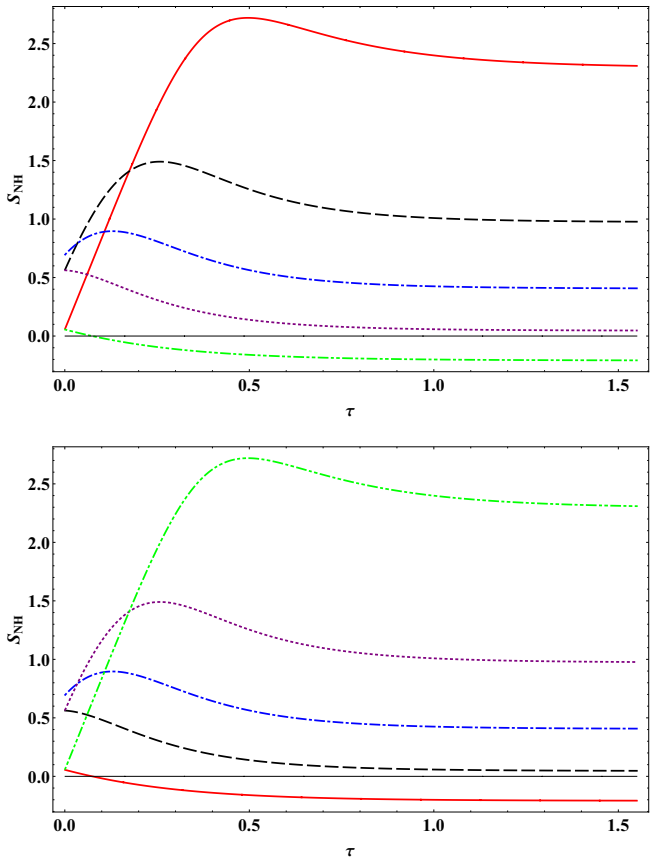,width=  0.99\columnwidth}\end{center}
\caption{Entropy $S_\text{NH}$ (\ref{e:m2entr2}) versus $\tau$
for the values of  $\tilde\gamma = -2$ (top plot)
and $\tilde\gamma =  2$ (bottom plot),
at the different values of $p$:
$0.01$ (solid line),
$1/4$ (dashed line),
$1/2 $ (dash-dotted line),
$3/4 $ (dotted line)
and
$0.99$ (dash-double-dotted line).}
\label{f-m2e2-gampm2}
\end{figure}

For the previous two-level model we found out that the NH entropy
goes to negative values during time evolution.
Here we illustrate that this behavior can be changed just
by adding a constant decay operator to the $\hat\Gamma$ operator, 
as explained in the last paragraphs of section \ref{s-ent}.

Thus, the Hermitian part of the model is given by
(\ref{eq:Hmod})
whereas the $\hat\Gamma$ operator is (in units $k_B = 1$):
\begin{equation}
\hat\Gamma
=
\hbar
\gamma 
\hat\sigma_z 
+
\hbar  \mu \hat I
.
\label{eq:Gammamod2}
\end{equation}
Unless otherwise specified, here and in the following, we assume
the notation of section \ref{s-tldet}.
The initial conditions for the evolution equation remain (\ref{eq:rhoini}).

Using the results given in the Appendix,
one can easily show that, for this model, both the normalized
density $\hat\rho$
and von Neumann entropy $S_\text{vN}$
are the same as those obtained for the model in Sec. \ref{s-tldet}.
However, the operator (\ref{e:m1omega}) acquires 
the factor $\exp{(- 2 \mu t)}$:
\be\lb{e:m2omega}
\hat\Omega 
=
\frac{1}{2 \mu^2}
\text{e}^{- 2 \mu t}
\left[
2 f_y (t) \hat\sigma_y
+
f_z (t) 
\hat\sigma_z
+
F(t)
\hat I
\right]
.
\ee
Therefore, using the transformations in Eq. (\ref{eq:gaugetrans}),
we obtain
\be
S_\text{NH}
=
S_{\rm vN}
-
\ln{\left[F (t)/\mu^2\right]}
+
2 \mu \tau
,
\lb{e:m2entr2}
\ee
where $S_{\rm vN}$ is given by (\ref{e:m1entr}).
The asymptotical value 
is  given by
\be
\lim\limits_{t\to +\infty} S_\text{NH} 
= 
\ln{
\left[
\frac{2 \mu^2}{\tilde\gamma (\tilde\gamma - \mu \bar p)}
\right]
} 
.
\ee
Unlike its analogue in Sec. \ref{s-tldet},
the $ S_\text{NH}$ entropy for this model is bounded.
As a matter of fact,
considering the more general form of the $\hat\Gamma$ operator,
$\hat\Gamma = \hbar \gamma \hat\sigma_z +
k \hbar  \mu \hat I$ (where $k$ is an arbitrary real number),
one can establish that the value $k=1$ 
acts as a critical threshold:
the NH
entropy decreases asymptotically
if $k<1$ (see the model in Sec. \ref{s-tldet}),
while it increases if $k>1$
(see the model in Sec. \ref{s-tldet3}).

\begin{figure}[htbt]
\begin{center}\epsfig{figure=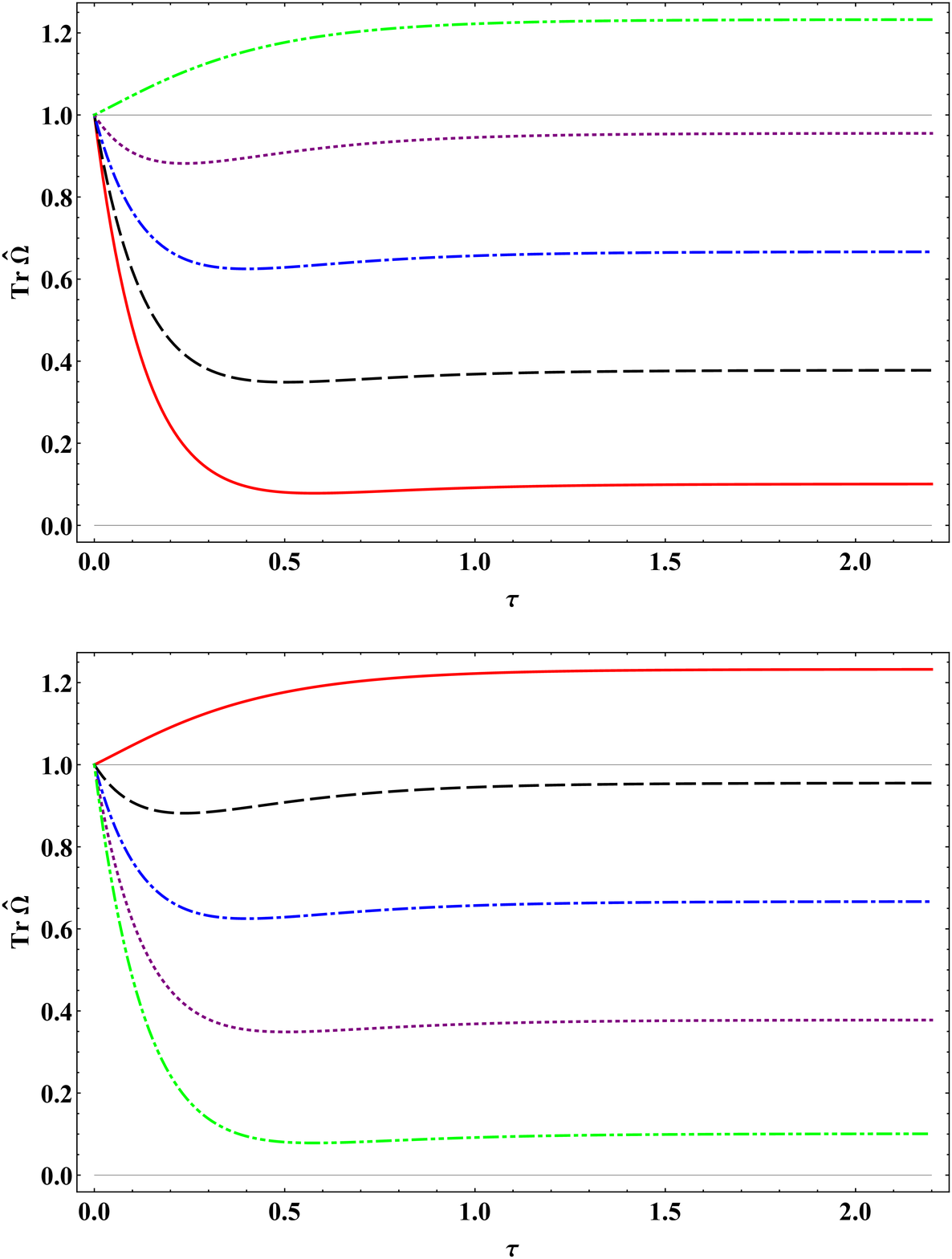,width=  0.99\columnwidth}\end{center}
\caption{Profiles of 
$\text{Tr}\, \hat\Omega$, where $\hat\Omega$ is given by (\ref{e:m2omega}), versus $\tau$,
for the values of  $\tilde\gamma = -2$ (top plot)
and $\tilde\gamma =  2$ (bottom plot),
at the different values of $p$:
$0.01$ (solid line),
$1/4$ (dashed line),
$1/2 $ (dash-dotted line),
$3/4 $ (dotted line)
and
$0.99$ (dash-double-dotted line).}
\label{f-m2tr-gampm2}
\end{figure}

The profiles of the von Neumann entropy, the NH entropy 
and the trace of the operator $\hat\Omega$
for this model
%for the initial state (\ref{eq:rhoini})
are shown in Figs. \ref{f-m1-gampm2}, \ref{f-m2e2-gampm2} and \ref{f-m2tr-gampm2}, respectively.
On can see that, 
as for the model in Sec. \ref{s-tldet},
the entropy $S_\text{NH}$
may takes negative values for a certain range of parameters
(when the trace of $\hat\Omega$ goes above one).
This will be discussed in details in the concluding section.

\sscn{Two-level tunneling model with non-Hermitian detuning and increasing NH entropy}{s-tldet3}

In the two previous cases, we found that the NH entropy
can assume a negative infinite value or a constant value.
Here we consider another model that, instead,
provides an entropy that asymptotically increases  with time.
Also in this case, such a behavior can achieved just
by adding a constant term to the $\hat\Gamma$ operator, 
as explained in the last paragraphs of section \ref{s-ent}.

\begin{figure}[htbt]
\begin{center}\epsfig{figure=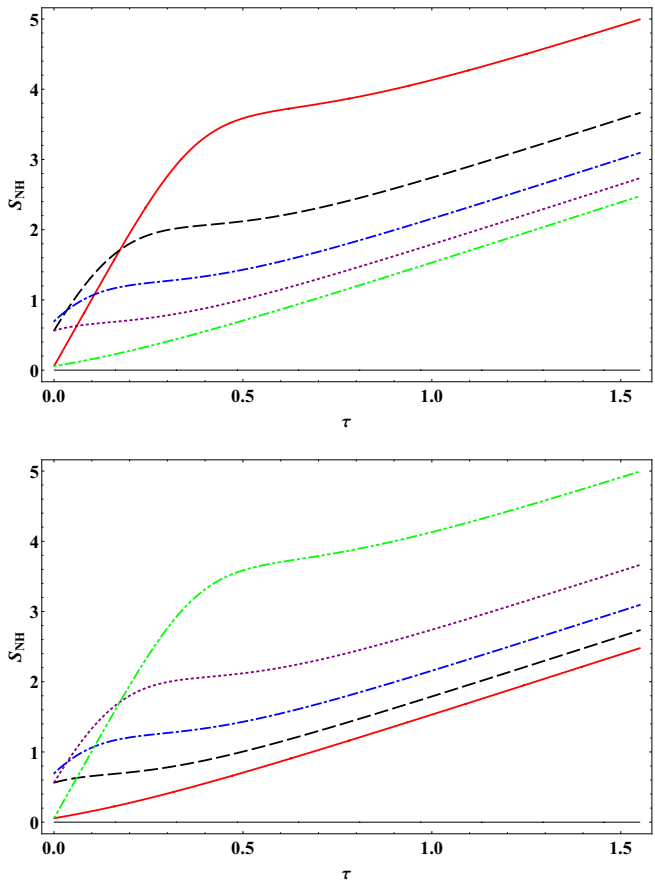,width=  0.99\columnwidth}\end{center}
\caption{Entropy $S_\text{NH}$ (\ref{e:m3entr2}) versus $\tau$
for the values of  $\tilde\gamma = -2$ (top plot)
and $\tilde\gamma =  2$ (bottom plot),
at the different values of $p$:
$0.01$ (solid line),
$1/4$ (dashed line),
$1/2 $ (dash-dotted line),
$3/4 $ (dotted line)
and
$0.99$ (dash-double-dotted line).}
\label{f-m3e2-gampm2}
\end{figure}

The Hermitian part of this model is still given
by Eq. (\ref{eq:Hmod})
whereas
the $\hat\Gamma$ operator is (in units $k_B = 1$):
\begin{equation}
\hat\Gamma
=
\hbar
\gamma 
\hat\sigma_z 
+
\frac{3}{2}
\hbar  \mu \hat I
.
\label{eq:Gammamod3}
\end{equation}
As one can see, the constant term multiplying the identity
operator in the definition of $\hat\Gamma$ above
is $\tfrac{3}{2} \hbar  \mu$.
Although, this case is qualitatively similar to 
any other one with the constant term's coefficient larger
than $\hbar  \mu$.
The initial conditions for the evolution equation remain (\ref{eq:rhoini}).

Using the results given in the Appendix,
one can easily show that for this model both the normalized
density $\hat\rho$
and von Neumann entropy $S_\text{vN}$
are the same as for the models in Secs. \ref{s-tldet}
and \ref{s-tldet2},
whereas the operator (\ref{e:m1omega}) acquires 
the factor $\exp{(- 3 \mu t)}$:
\be\lb{e:m3omega}
\hat\Omega 
=
\frac{1}{2 \mu^2}
\text{e}^{- 3 \mu t}
\left[
2 f_y (t) \hat\sigma_y
+
f_z (t) 
\hat\sigma_z
+
F(t)
\hat I
\right]
.
\ee
Therefore, using (\ref{eq:gaugetrans}),
we obtain
\be
S_\text{NH}
=
S_{\rm vN}
-
\ln{\left[F (t)/\mu^2\right]}
+
3 \mu \tau
,
\lb{e:m3entr2}
\ee
where $S_{\rm vN}$ is given by (\ref{e:m1entr}).
The asymptotical value 
is  given by
\be
\lim\limits_{t\to +\infty} S_\text{NH} 
=
\ln{\mu^2} - \lim\limits_{t\to\infty} \ln{F (t)} + 3 \mu \tau
\propto \mu \tau
,
\ee
such that $S_\text{NH}$ tends to a linear function with a positive coefficient
at large times.

\begin{figure}[htbt]
\begin{center}\epsfig{figure=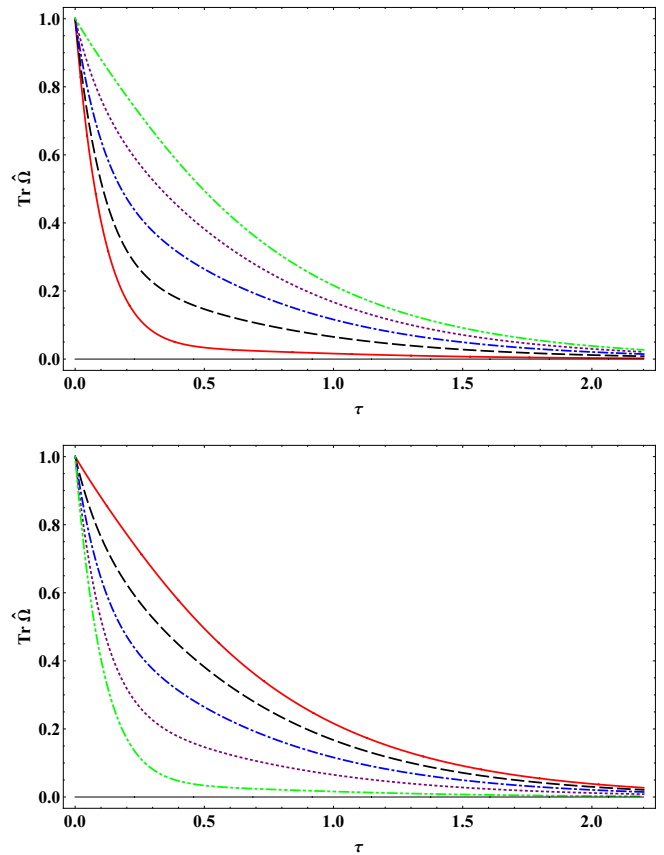,width=  0.99\columnwidth}\end{center}
\caption{Profiles of 
$\text{Tr}\, \hat\Omega$, where $\hat\Omega$ is given by (\ref{e:m3omega}), versus $\tau$,
for the values of  $\tilde\gamma = -2$ (top plot)
and $\tilde\gamma =  2$ (bottom plot),
at the different values of $p$:
$0.01$ (solid line),
$1/4$ (dashed line),
$1/2 $ (dash-dotted line),
$3/4 $ (dotted line)
and
$0.99$ (dash-double-dotted line).}
\label{f-m3tr-gampm2}
\end{figure}

The profiles of the von Neumann entropy, NH entropy 
and the trace of the operator $\hat\Omega$
for this model
%for the initial state (\ref{eq:rhoini})
are shown in Figs. \ref{f-m1-gampm2}, \ref{f-m3e2-gampm2} and \ref{f-m3tr-gampm2}, respectively.

\scn{Discussion and conclusions}{s-con}

In this paper we have provided a generalized formulation
of the quantum fine-grained entropy for systems described by
non-Hermitian Hamiltonians.
We have adopted a straightforward generalization of the von Neumann entropy,
defined in terms of the normalized density matrix (obeying a nonlinear 
equation of motion),
and introduced another definition of an entropy, $S_{\rm NH}$,
in terms of the normalized average
of the logarithm of the non-normalized density matrix.
We have shown that in both cases the entropy production is non-zero.
However, 
%upon explicitly studying few non-Hermitian models
we have found that it is 
$S_{\rm NH}$ that
properly captures the physical behavior of the probability and disorder in a system
in presence of sinks or sources described by non-Hermitian Hamiltonians.

In Sec. \ref{s-ex} we have studied some models in order
to illustrate the different behavior of the 
$S_{\rm vN}$ and $S_{\rm NH}$ entropies.
In particular, for the models in Secs. \ref{s-tldet}-\ref{s-tldet3}
both the normalized density operator, $\hat\rho$,
and the von Neumann entropy, $S_\text{vN}$, do not change
while the entropy $S_{\rm NH}$ does.
We show that at large times the value $S_{\rm NH}$
decreases asymptotically 
for the model of Sec. \ref{s-tldet},
it tends to a finite value 
for the model in Sec. \ref{s-tldet2}, and
it increases for the model in Sec. \ref{s-tldet3}.

%\rd{TO BE REVISED:
%The explanation of what occurs in the two-level model we have studied can unfold
%as follows.
%If the system is isolated then its dynamics is governed by the Hermitian Hamiltonian.
%In the presence of sinks or sources, states can become metastable,
%therefore, the probability is not necessarily preserved.
%The non-Hermitian approach deals with such cases in terms
%of the decay operator $\hat{\Gamma}$ (\emph{i.e.}, the anti-Hermitian part
%of the Hamiltonian).
%This operator effectively represents the environment,
%hence, the dimensionality of the system's Hilbert space stays unchanged but
%the information about the probability flow from/to the 
%reservoir can be obtained from the entropy $S_\text{NH}$
%and the trace of the density operator $\hat\Omega$.
%We can have the following situations:
%(i) both levels are occupied, hence the entropy must be maximum;
%(ii) only one level is occupied, hence the entropy must be zero;
%(iii) only one level is occupied with a leakage toward the
%reservoir, hence the entropy must be negative.
%The case (iii) can be monitored through  $\text{Tr}\, \hat\Omega$.
%When there is an inflow of probability from the reservoir,
%both $\text{Tr}\, \hat\Omega$ and the entropy should grow.
%This cannot be observed by using the conventional entropy $S_{\rm vN}$ 
%(cf. Fig. \ref{f-m1-gampm2}).
%However, these processes can be visualized
%by using the entropy $S_\text{NH}$, as one can see from
%the presented figures.
%}
The results of the present work,
when considered together
with our previous
studies \cite{sz14,ks,sz14cor}, allow us to draw a certain number
of conclusions.
Non-Hermitian dynamics
is able to describe in the quantum realm the production of entropy
in a way similar to what non-Hamiltonian dynamics with phase space
compressibility does in the classical realm.
Non-Hermitian dynamics also seems to need both the
normalized density operator, $\hat{\rho}$,
and non-normalized one, $\hat{\Omega}$,
in order to provide a proper statistical theory.
The non-normalized density operator, defined as a solution of Eq. (\ref{eq:dotOmega}),
captures some important features of the decay process, 
such as
the non-conservation of probability in the (sub)system and its
``leakage'' into the surrounding environment.
The normalized density operator guarantees that the
probabilistic interpretation of averages can be maintained.
In this regard, the entropy $S_\text{NH}$ combines both
operators in a proper way and can signal
the expected thermodynamic behavior
of an open system.
The entropy $S_\text{NH}$ also
seems more suitable for describing the gain-loss processes that
are related to the probability's non-conservation, since
it contains information not only about
the conventional von Neumann entropy 
$S_\text{vN}$ 
but also about the trace of the operator 
$\hat\Omega$,
according to the relation (\ref{e:diffentr}).
Assuming that $S_\text{vN}$ is bound at large times,
the NH entropy grows when $\text{Tr}\,\hat\Omega$
decreases, also it takes positive values if $\text{Tr}\,\hat\Omega < 1$
and negative ones otherwise.
Hence,
one can say that $S_\text{NH}$ describes
the flow of information between the system and the bath.

Further studies are needed to understand whether and when
$S_{\rm NH}$ may deserve a complete quantitative thermodynamic status
as well as whether there might be viable physical interpretations
of a negative entropy
(not necessarily given in terms of the number
of occupied microscopic states).

\section*{Acknowledgments}

This research was supported by
the National Research Foundation of South Africa
under Grant 
%95965 and 
98892.

\appendix*
\scn{Hamiltonian shift transformations and entropy}{app} 

Following the discussions presented in Refs. \cite{sz14,ks}, let us
consider the following transformation 
of the $\hat\Gamma$ operator
\be\lb{eq:transgfau}
\hat\Gamma = \hat\Gamma' + \frac{1}{2}\hbar \alpha \hat I
,
\ee
where $\alpha$ is an arbitrary real constant and
$\hat I$ is the unity operator.
This transformation is a subset of the transformation
\be
\hat{\cal H} = \hat{\cal H}' + c_0 \hat I, 
\label{eq:H-const-shift}
\ee
$c_0$ being an arbitrary complex number, which is the non-Hermitian generalization
of the energy shift in conventional quantum mechanics.
Therefore, in Refs. \cite{sz14,ks} it was called
the ``gauge'' transformation of the Hamiltonian (\ref{eq:tot-NH-H}),
whereas the terms of the type $c_0 \hat I$ can be called the ``gauge'' terms.

In Ref. \cite{ks} it was shown that the equation (\ref{eq:dotrho}) is invariant under the
transformation (\ref{eq:transgfau}),
therefore, one immediately obtains
\be
\hat\rho = \hat\rho', \
S_\text{vN} = S_\text{vN}'
,
\ee
therefore the von Neumann entropy is not affected
by the transformation 
%in Eq. 
(\ref{eq:transgfau}).
One can see that any information regarding the
shifting of the total non-Hermitian Hamiltonian 
is lost if one deals solely with the normalized density operator.

However, the evolution equation (\ref{eq:dotTrOmega}) is not
invariant under the shift (\ref{eq:transgfau}).
If $\hat{\cal H}$ is time-independent then,
substituting (\ref{eq:transgfau}) into (\ref{eq:dotTrOmega}),
we obtain that the non-normalized density acquires an exponential 
factor:
\be
\hat\Omega = \hat\Omega' \exp{(-\alpha t)},
\ee
therefore, recalling the relation (\ref{e:diffentr}), we obtain
\be
S_\text{NH} 
= 
S_\text{vN}' - k_B \ln{\text{Tr}\, \hat\Omega}
=
S_\text{NH}' + k_B \alpha t
,
\ee
which indicates that any lost information 
about the shifting term in Eq. (\ref{eq:H-const-shift})
in the total non-Hermitian Hamiltonian, due the normalization
procedure in Eq. (\ref{eq:rho}),
can be recovered by means of the NH entropy.

%\newpage


\begin{thebibliography}{}
%1
\bibitem{callen}
H. B. Callen, 
Thermodynamics and an Introduction to Thermostatistics
(John Wiley \& Sons, New York, 1985).
%2
\bibitem{aharonov}
Y. Aharonov and D. Rohrlich, Quantum Paradoxes
(Wiley-VCH, Weinheim, 2005).
%3
\bibitem{balescu}
R. Balescu, Equilibrium and Nonequilibrium Statistical Mechanics
(John Wiley \& Sons, New York, 1975).
%4
\bibitem{nose}
S. Nos\`e, Mol. Phys. {\bf 52}, 255 (1984).
%5
\bibitem{hoover}
W. G. Hoover, Phys. Rev. A {\bf 31}, 1695 (1985).
%6
\bibitem{b}
A. Sergi and M. Ferrario,
%Non-Hamiltonian  Equations of Motion with a Conserved Energy,
Phys. Rev. E {\bf 64}, 056125 (2001).
%7
\bibitem{b2}
A. Sergi,
%Non-Hamiltonian Equilibrium Statistical Mechanics,
Phys. Rev. E {\bf 67}, 021101 (2003).
%8
\bibitem{aspvg}
A. Sergi and P. V. Giaquinta,
%On the geometry and entropy of non-Hamiltonian phase space,
J. Stat. Mech. Theory and Exp. {\bf 02}, P02013 (2007).
%9
\bibitem{andrey}
L. Andrey,
%The rate of entropy change in non-Hamiltonian systems,
Phys. Lett. A {\bf 111}, 45 (1985).
%10
\bibitem{andrey2}
L. Andrey,
%Note concerning the paper "The rate of entropy change in non-Hamiltonian systems",
Phys. Lett. A {\bf 114}, 183 (1986).
%11
\bibitem{optics}
C. E. R\"uter, K. G. Makris, R. El-Ganainy, D. N. Christodoulides,
M. Segev, and D. Kip, Nature Phys. {\bf 6}, 192 (2010).
%12
\bibitem{optics2}
A. Guo, G. J. Salamo, D. Duchesne, R. Morandotti, M. Volatier-Ravat, V. Aimez, G. A. Siviloglou, and D. N. Christodoulides,
Phys. Rev. Lett. {\bf 103}, 093902 (2009).
%13
\bibitem{nimrod2}
N. Moiseyev, Phys. Rep. {\bf 302}, 211 (1998).
%14
\bibitem{seba}
W. John, B. Milek, H. Schanz, and P. Seba,
Phys. Rev. Lett. {\bf 67}, 1949 (1991).
%15
\bibitem{spyros}
C. A. Nicolaides and S. I. Themelis,
Phys. Rev. A {\bf 45}, 349 (1992).
%16
\bibitem{fesh}
H. Feshbach, Ann. Phys. {\bf 5}, 357 (1958).
%17
\bibitem{fesh2}
H. Feshbach, Ann. Phys. {\bf 19}, 287 (1962).
%18
\bibitem{sudarshan}
E. C. G. Sudarshan, Phys. Rev. D {\bf 18}, 2914 (1978).
%19
\bibitem{selsto}
S. Selst\o, T. Birkeland, S. Kvaal, R. Nepstad, and M. F\o rre,
J. Phys. B: At. Mol. Opt. Phys. {\bf 44}, 215003 (2011).
%20
\bibitem{baker}
H. C. Baker, Phys. Rev. Lett. {\bf 50}, 1579 (1983).
%21
\bibitem{baker2}
H. C. Baker,
Phys. Rev. A {\bf 30}, 773 (1984).
%22
\bibitem{chu}
S.-I. Chu and W. P. Reinhardt, Phys. Rev. Lett. {\bf 39}, 1195 (1977).
%23
\bibitem{kor64}
J. Korringa,
%Dynamical Decomposition of a Large System
Phys. Rev. {\bf 133}, 1228 (1964).
%24
\bibitem{wong67}
J. Wong, J. Math. Phys. {\bf 8}, 2039 (1967).
%25
\bibitem{heg93}
G. C. Hegerfeldt,
Phys. Rev. A {\bf 47}, 449 (1993).
%26
\bibitem{bas93}
S. Baskoutas, A. Jannussis, R. Mignani, and V. Papatheou,
J. Phys. A {\bf 26}, L819 (1993).
%27
\bibitem{ang95}
P. Angelopoulou, S. Baskoutas, A. Jannussis, R. Mignani, and V. Papatheou,
Int. J. Mod. Phys. B {\bf 9}, 2083 (1995).
%28
\bibitem{rotter}
I. Rotter, arXiv:0711.2926.
%29
\bibitem{rotter2}
I. Rotter, J. Phys. A {\bf 42}, 153001 (2009).
%30
\bibitem{gsz08}
H. B. Geyer, F. G. Scholtz and K. G. Zloshchastiev,
in: Proceedings of $12^{\rm th}$ International Conference on
Mathematical Methods in Electromagnetic Theory
(Odessa, 2008) pp. 250-252.
%31
\bibitem{bellomo}
R. Lo Franco, B. Bellomo, S. Maniscalco, and G. Compagno,
% Dynamics of quantum correlations in two-qubit systems
%with non-Markovian environments,
Int. J. Mod. Phys. B {\bf 27}, 1345053 (2013).
%32
\bibitem{banerjee}
S. Banerjee and R. Srikanth,
% Complementarity in generic open quantum systems,
Mod. Phys. Lett. B {\bf 24}, 2485 (2010).
%33
\bibitem{reiter}
F. Reiter and A. S. S\o rensen,
%Effective operator formalism for open quantum systems
Phys. Rev. A {\bf 85}, 032111 (2012).
%34
\bibitem{bg12}
D. C. Brody and E.-M. Graefe,
Phys. Rev. Lett. {\bf 109}, 230405 (2012).
%35
\bibitem{sz14}
K. G. Zloshchastiev and A. Sergi,
J. Mod. Optics {\bf 61}, 1298 (2014).
%36
\bibitem{ks}
A. Sergi and K. Zloshchastiev, 
%Non-Hermitian quantum dynamics of a two-level system and models of
%dissipative environments,
Int. J. Mod. Phys. B {\bf 27}, 1350163 (2013).
%37
\bibitem{vonneumann}
J. von Neumann, 
The Mathemathical Foundation of Quantum Mechanics
(Princeton University Press, Princeton, 1955).

\bibitem{bpbook}
H.-P. Breuer and F. Petruccione, \textit{The Theory
of Open Quantum Systems} (Oxford University Press, 2002). 

%38
\bibitem{emg-korsch}
E.-M. Graefe, M. H\"oning, and H. J. Korsch,
%Classical limit of non-Hermitian quantum
%dynamics—a generalized canonical structure,
J. Phys. A {\bf 43}, 075306 (2010).
%39
\bibitem{zno02}
M. Znojil,
%Should PT Symmetric Quantum Mechanics Be Interpreted as Nonlinear?
J. Nonlin. Math. Phys. \textbf{9}, 122-133 (2002).
%40
\bibitem{sz14cor}
A. Sergi and K. Zloshchastiev, 
%Time correlation functions for non-Hermitian quantum systems,
Phys. Rev. A {\bf 91}, 062108 (2015).
%41
\bibitem{oerter}
R. N. Oerter, 
%Entropy and irrversibility in the quantum realm,
Am. J. Phys. {\bf 79}, 297 (2011).
%42
\bibitem{balian}
R. Balian,
%Incomplete descriptions and relevant entropies,
Am. J. Phys. {\bf 67}, 1078 (1999).
%43
\bibitem{leggett}
A. J. Leggett, \textit{et al.},
%S. Chakravarty, A. T. Dorsey, M. P. A. Fisher, A. Garg, and M. Zwerger, 
Rev. Mod. Phys. {\bf 59}, 1 (1987).
%44
\bibitem{sgh92}
F. G. Scholtz, H. B. Geyer and F. J. W. Hahne,
%Quasi-Hermitian operators in quantum mechanics and the variational principle}{
Ann. Phys. {\bf 213}, 74 (1992).
%45
\bibitem{ben98} 
C. M. Bender and S. Boettcher,
%Real spectra in non-Hermitian Hamiltonians having PT symmetry}{
Phys. Rev. Lett. {\bf 80}, 5243 (1998).


\end{thebibliography}
\end{document}